# Preventing Aluminum Photocorrosion for Ultraviolet Plasmonics

Aleksandr Barulin, Jean-Benoît Claude, Satyajit Patra, Antonin Moreau, Julien Lumeau, Jérôme Wenger*

*Aix Marseille Univ, CNRS, Centrale Marseille, Institut Fresnel, 13013 Marseille, France*

\* Corresponding author: jerome.wenger@fresnel.fr

**Abstract**

Ultraviolet (UV) plasmonics aims at combining the strong absorption bands of molecules in the UV range with the intense electromagnetic fields of plasmonic nanostructures to promote surface-enhanced spectroscopy and catalysis. Currently, aluminum is the most widely used metal for UV plasmonics, and is generally assumed to be remarkably stable thanks to its natural alumina layer passivating the metal surface. However, we find here that under 266 nm UV illumination, aluminum can undergo a dramatic photocorrosion in water within a few tens of seconds and even at low average UV powers. This aluminum instability in water environments critically limits the UV plasmonics applications. We show that the aluminum photocorrosion is related to the nonlinear absorption by water in the UV range leading to the production of hydroxyl radicals. Different corrosion protection approaches are tested using scavengers for reactive oxygen species and polymer layers deposited on top of the aluminum structures. Using optimized protection, we achieve a ten-fold increase in the available UV power range leading to no visible photocorrosion effects. This technique is crucial to achieve stable use of aluminum nanostructures for UV plasmonics in aqueous solutions.

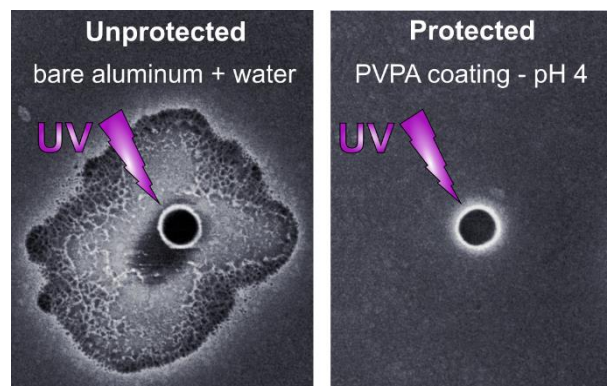

Figure for Table of Contents

**Introduction**

Plasmonics offers outstanding possibilities to create intense local electric fields, which can benefit to many light-driven applications including sensing,[1] photoemission,[2] light harvesting,[3] photodetection,[4] or catalysis.[5] As most organic molecules feature strong absorption bands in the UV spectral domain, extending plasmonics into the 200-400 nm ultraviolet range is of major interest to further promote sensing and catalysis applications.[6–9] However, gold and silver, the classical metals used for plasmonics in the visible and near-infrared spectral ranges, fail to operate in the UV regime due to their strong losses and interband transitions below 400 nm. Currently aluminum is the most widely used metal for UV plasmonics,[10,11] owing to its good optical properties down to 200 nm, low cost and CMOS compatibility.[12–14] In the visible range, aluminum plasmonics covers a wide range of research, including surface enhanced Raman scattering,[15,16] spectroscopy,[17–20] fluorescence sensing,[21–24] strong coupling,[25] photodetection,[26,27] photovoltaics,[28] photocatalysis,[29,30] water desalination,[31] and color filters.[32–34] However, while many theoretical works have outlined the high potential of aluminum for UV plasmonics,[13,35–38] the aluminum applications operating in the 200-400 nm UV range remain scarce, and are often limited to evaporated samples or non-aqueous solvents.[7–9,39–43]

A largely overlooked issue limiting the use of aluminum for plasmonics is corrosion in water environment.[16,30] While aluminum is quite stable in air due to its natural oxide layer, it can corrode when exposed to water medium.[44–46] In this case, the water molecules can induce pitting corrosion at the junction between the metal grains where the oxide layer is thinner or has cracks. In the dark and in the absence of chloride ions, the corrosion of aluminum layers by water remains quite slow,[16] and several hours of exposition to water are needed in order to yield visible effects. However, we have found that the situation is strikingly different when ultraviolet light is present.

Here, we use a generic platform to investigate the stability of aluminum nanostructures for UV plasmonics in a water environment. A 266 nm laser beam is focused on single nanoapertures milled in aluminum and filled with different solutions. While the optical energy per pulse is kept low enough to avoid any direct photodamage of the metal layer, the presence of water molecules can lead to a dramatic UV photocorrosion of aluminum within only a few seconds. We investigate the origin of this effect and relate it to the two-photon UV absorption of water producing hydroxyl radicals. Scavengers for reactive oxygen species in solution improve the aluminum stability and mitigate the photocorrosion effects. Additionally, passivating the aluminum surface with polyvinylphosphonic acid (PVPA)[47] and polydopamine (PDA)[48] layers further prevents the photocorrosion and significantly extends the accessible UV power range by ten-fold. Keeping the aluminum nanostructures stable under intense illumination is crucial for every UV plasmonics application. Improving our understanding of aluminum photocorrosion in water and developing appropriate protection strategies are therefore important steps to enable UV plasmonic sensing and catalysis applications in aqueous solutions.

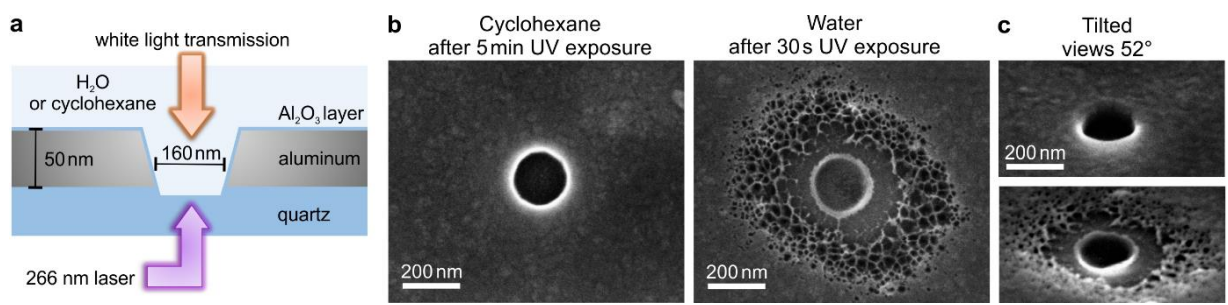

**Figure 1.** Photocorrosion of aluminum nanoapertures by ultraviolet light. (a) Experimental scheme of a single nanoaperture milled in an opaque aluminum film illuminated by a focused ultraviolet laser beam. The aperture and the upper medium are filled with pure water or cyclohexane. The white light transmission serves to monitor *in situ* the increase of the apparent aperture diameter during the photocorrosion process. (b) Representative scanning electron microscopy images of a nanoaperture of 160 nm diameter filled with cyclohexane or pure water. The average UV power for both cases is 200 µW, corresponding to a peak intensity of 90 mW/µm² focused on the nanoaperture. (c) Tilted views of the nanoapertures in (b) where the sample has been tilted by 52°.

**Results and Discussion**

In this work we study nanoapertures (also sometimes called zero-mode waveguides[21]) of diameters ranging from 60 to 210 nm milled in a 50 nm thick aluminum layer on a quartz coverslip. Nanoapertures feature distinctive advantages to benchmark the influence of UV light on aluminum corrosion: the transmission through this sub-wavelength diameter is highly sensitive to the aperture size and can be monitored *in situ* during the photocorrosion process, the opaque 50 nm metal layer allows to work on a dark background and collect only the optical signal stemming from the aperture, and the nanoapertures can be easily and reproducibly fabricated by focused ion beam (FIB) milling.[49,50] Figure 1a shows a scheme of our setup. Briefly, a 266 nm laser beam is focused onto a single aluminum nanoaperture by a 0.6 NA UV microscope objective, leading to a laser spot size at the focus of 250 nm full width at half maximum (FWHM). A confocal photomultiplier detector conjugated to the laser focus spot records the transmission of the white light (detection range 310-410 nm) through the aperture.

When the aperture is filled with cyclohexane, no photodamage is visible on scanning electron microscopy (SEM) images, even after prolonged exposure to 200 µW focused UV light (Fig. 1b). This importantly shows that the energy fluence per pulse (equivalent to 0.6 mJ/cm²) is well below the aluminum photodamage threshold by UV light and that the laser pulses alone do not affect the nanoaperture structure.[51] However, when the aperture is filled with water instead of cyclohexane, the same experiment leads to a dramatic photocorrosion of aluminum in a very short time of only a few seconds (Fig. 1b,c). The SEM images show that the aluminum around the nanoaperture has been almost completely dissolved (only the undercut in the quartz substrate remains) and the interface between the aluminum layer and the exposed area now features a very porous region with pittings of diameters comparable to the aluminum grain sizes.

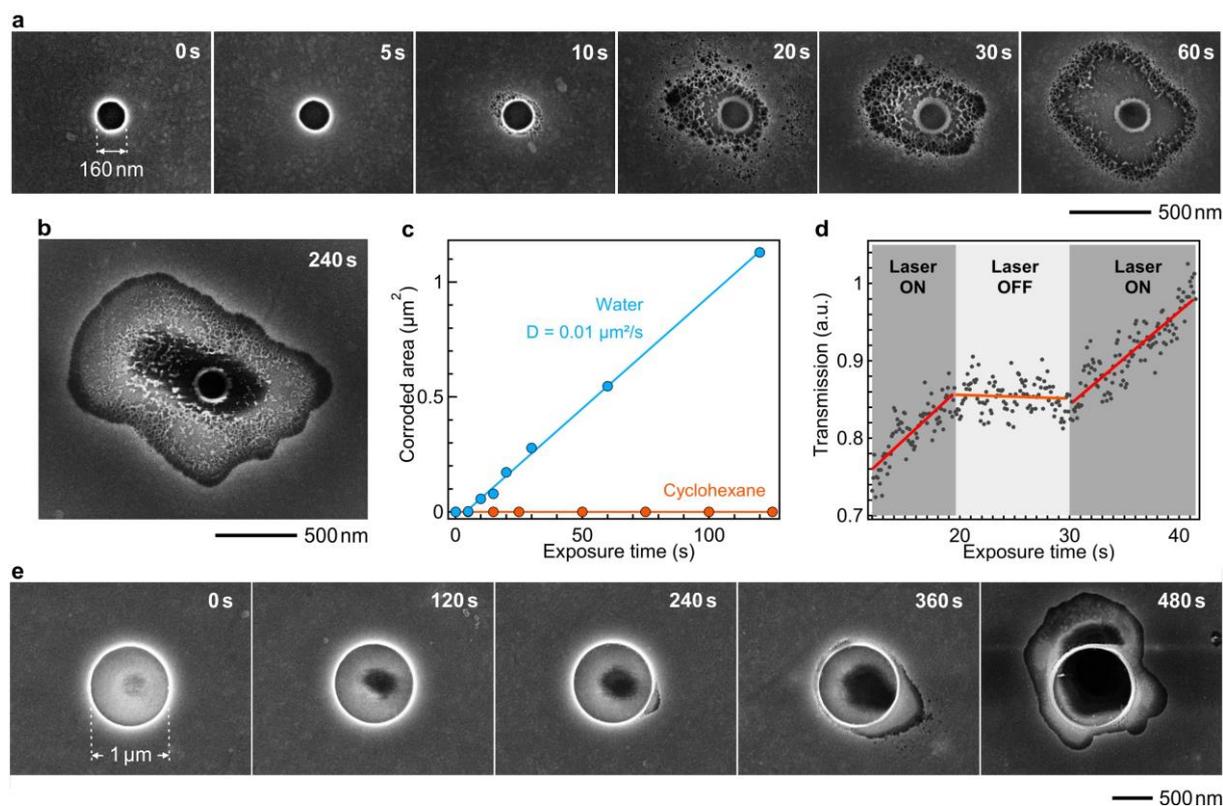

**Figure 2.** Temporal evolution of the UV photocorrosion of aluminum apertures filled with pure water. (a,b) Scanning electron microscopy (SEM) images of 160 nm diameter apertures filled with pure water (pH 7) taken after increasing irradiation times with 200 µW average UV power. (c) Evolution of the corroded aluminum area as a function of the irradiation time. No corrosion was detected when the aperture is filled with cyclohexane. (d) Temporal evolution of the white light transmission through the nanoaperture. The UV laser is switched off between 20 and 30 s, leading to no additional photocorrosion during this time window. (e) Same experiment as (a) for a 1 µm diameter aperture. In this case, no aluminum is directly illuminated by the UV light as the aperture diameter is twice bigger than the UV laser spot diameter.

Figure 2 investigates the temporal dynamics of the water-induced photocorrosion of aluminum. Different similar nanoapertures of 160 nm diameter are illuminated with increasing exposure times (Fig. 2a,b), allowing us to extract the evolution of the corroded area versus the exposure time (Fig. 2c). We find that the corroded area scales linearly with the exposure time (as expected for a diffusion-limited process), with a rate of 0.01 µm²/s. Figure 2a also shows that the corrosion is initiated very quickly within 5 to 10 seconds once the aperture is exposed to UV. To further confirm that the photocorrosion is controlled by UV light, we record the time evolution of the transmission through the nanoaperture (Fig. 2d). When the laser in on, the transmission grows with time, indicating that the aperture gets enlarged by corrosion. When the laser is blocked, the transmission remains at a constant level, consistent with the aperture keeping the same diameter. Finally when the laser illuminates again the aperture, the corrosion process restarts to increase the apparent aperture diameter and the optical transmission.

A first idea about the physical phenomenon responsible for the aluminum UV photocorrosion is that hot electrons from the aluminum are excited and extracted by the UV pulse.[5,52–54] These hot electrons then interact with water molecules to generate radicals that in turn corrode the aluminum layer.[55] In this view, the process should stop when no more aluminum is illuminated by the UV light to extract hot electrons. However, we observe corrosion areas larger than 1 µm² (Fig. 2b) which are much bigger than the UV spot size, invalidating the hot electron contribution from the aluminum. To clearly investigate this hypothesis, we study apertures of 1 µm diameter, 4 times bigger than the UV spot size FWHM. With this large diameter, no aluminum is directly illuminated by the UV light. Still, our experiments observe photocorrosion when the water filling the aperture is illuminated by UV (Fig. 2e). The illumination time needed to start seeing the photocorrosion effect increases, but this goes with the rate indicated on Fig. 2c: at 0.01 µm²/s, one has to wait at least 100 s to reach a 1 µm² area. Altogether, while the experiments in Fig. 1 and 2 show that water and UV are needed for the corrosion process, Fig. 2e shows that hot electrons stemming from the aluminum itself do not play a significant role here.

While there have been numerous studies on water photolysis by UV light, most focus on the photochemical processes in presence of a semiconductor or chlorine,[56–59] which are both absent in our case. The first absorption band of water occurs for wavelengths below 190 nm, or photon energies higher than 6.5 eV.[60] This is well higher than the 4.7 eV energy of the 266 nm photons and water should be transparent to this wavelength. However, in the case of pulsed picosecond illumination, water can absorb 266 nm light by a two-photon mechanism.[60,61] A laser peak intensity higher than $10^{12}$ W/m² was considered necessary for the two-photon absorption to be detected.[60] This level is about 10 times higher than in our 90 mW/µm² case, but the tight focusing of the UV laser in our experiment and the sensitive transmission detection may explain this difference. After absorbing two photons at 266 nm, a water molecule acquires an energy of 9.3 eV and can undergo photolysis by two mechanisms: ionization and dissociation.[60–62] In the ionization process, an electron is ejected:

$$H_2O^* \rightarrow H_2O^+ + e^- \qquad (1)$$

The cation radical of water is unstable and will further react with another water molecule to generate a hydronium ion and a hydroxyl radical:[60]

$$H_2O^+ + H_2O \rightarrow H_3O^+ + OH^\bullet \qquad (2)$$

Alternatively, the excited water molecule can also dissociate to generate a hydrogen atom and a hydroxyl radical:

$$H_2O^* \rightarrow H + OH^\bullet \qquad (3)$$

The hydrogen atom is expected to quickly collide with the oxygen atom from another water molecule to generate a hydronium ion and a solvated electron:[61]

$$H + H_2O \rightarrow H_3O^+ + e^- \qquad (4)$$

The probabilities for an excited water molecule to undergo photolysis were estimated to be 15% for ionization and 13% for dissociation,[60] hence both processes are as likely to occur. Following the primary reactions (1-4), hydroxyl radicals $OH^\bullet$, solvated electrons and hydronium ions are generated. These species will further react with each other and with oxygen dissolved in water, and importantly also with aluminum which will be dissolved into $Al^{3+}$ ions after reacting with hydroxyl radicals:

$$Al + 3\,OH^\bullet \rightarrow Al^{3+} + 3\,OH^- \qquad (5)$$

Alternatively, two hydroxyl radicals can combine to form one hydrogen peroxide molecule, which will also dissolve the aluminum:[63]

$$Al + 3\,H_2O_2 \rightarrow Al^{3+} + 3\,OH^\bullet + 3\,OH^- \qquad (6)$$

With the two processes (5,6), the corrosion of aluminum is mediated by hydroxyl radicals. Therefore, it should be possible to mitigate the corrosion effects by adding scavengers for reactive oxygen species to the buffer, and this is what we will investigate in the following.

According to the reaction (5), three hydroxyl radicals are required to fully dissolve one aluminum atom. Assuming that the reaction rate scales with the stoichiometric coefficient, the dissociation rate for aluminum should depend to the hydroxyl radical concentration by a power 3, and thus to the optical power by a $6^{th}$ power, as the two-photon generation of $OH^\bullet$ depends quadratically on the UV power. In the case of reaction (6), the dependence would be even of higher order. We measure the power dependence by performing a series of transmission measurements and gradually increasing the UV power until we start monitoring a transmission gain indicating photocorrosion (Fig. 3a). Importantly, each data point is taken after the same integration time so that the datasets can be readily compared between each other.

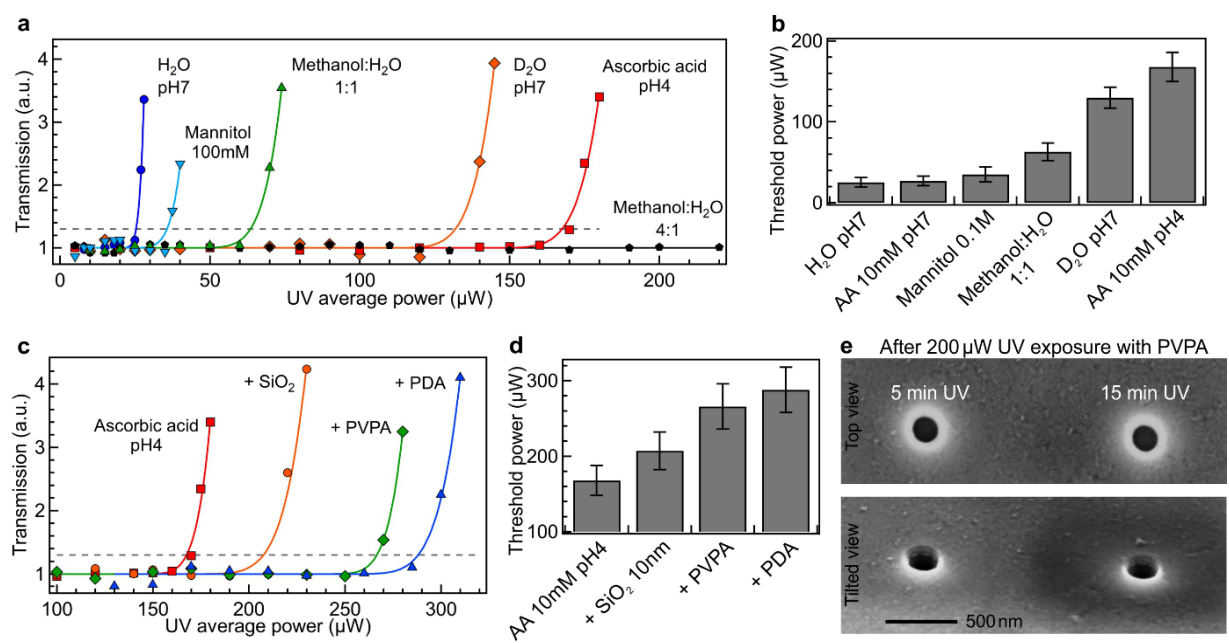

**Figure 3.** Photocorrosion inhibition with different scavengers and additional protection layers covering the aluminum surface. (a) Transmitted white light intensity through a 160 nm diameter aluminum aperture for increasing UV average powers with different solutions filling the aperture. Each data point is taken after a 2 min integration time, and lines are numerical fits with a $6^{th}$ power law. The power for which the transmission increases by more than 30% (gray dashed line) defines the threshold power. (b) Photocorrosion threshold power for different solutions filling the nanoaperture. AA stands for ascorbic acid. (c) Transmitted white light intensity through a 160 nm diameter aluminum aperture for increasing UV average powers, with 2 min integration time per point. The aluminum/alumina surface is left untreated (red curve, same as in Fig. 3a) or coated with 10 nm silica, polyvinylphosphonic acid (PVPA) or polydopamine (PDA). For all these experiments, the aperture is filled with 10 mM ascorbic acid (AA) at pH 4. (d) Photocorrosion threshold power with different additional protection layers. (e) SEM images of 160 nm diameter apertures coated with PVPA and filled with 10 mM AA (pH 4), after 5 or 15 minutes exposure at 200 μW average UV power. No photocorrosion is visible in these cases, while an untreated aperture filled with pure water would look like the image in Fig. 2b.

Figure 3a shows the transmission recorded while increasing the UV laser power for different solutions filling the aperture. The dependence with the laser power is clearly non-linear. For all the different buffer solutions tested in Fig. 3a, we find that the transmission follows a 6$^{th}$ power law with the laser intensity, as predicted from the stoichiometric dependence with the hydroxyl radical concentration in reaction (5). We use these datasets to record for each diameter a threshold power, defined as the optical power for which the transmission increases by 30% above the initial transmission before the photocorrosion started (Fig. 3b). The choice of 30% increase is quite arbitrary and is chosen to ensure that we are well above the noise level of our measurements.

The results in Fig. 3a,b indicate that the composition of the medium filling the aperture plays a key role in inhibiting the photocorrosion. Pure water at pH 7 is found to have the lowest power threshold, for which photocorrosion is most likely to occur at moderate optical powers. Mannitol and methanol, both known scavengers for OH• hydroxyl radicals,[64,65] displace the occurrence of photocorrosion towards higher optical powers. This corrosion inhibition effect confirms the central role played by hydroxyl radicals in the photocorrosion of aluminum. Conversely, oxygen scavengers such as sodium azide or glucose oxidase[66] do not change significantly the behavior as compared to pure water (data not shown). Ascorbic acid is a good reducing agent,[67] but at pH 7 it does not influence significantly the photocorrosion of aluminum. However, at pH 4, ascorbic acid efficiently inhibits the photocorrosion (Fig. 3a,b). This is largely related to decreasing the pH, as similar results could be obtained by replacing ascorbic acid by acetic acid. The pH influence further confirms the role of hydroxyl radicals in the photocorrosion. However, it is not possible to further reduce the pH, as the aluminum and its alumina protective layer are no more stable below pH 4.[45] Interestingly, using deuterated (heavy) water D$_2$O, where each hydrogen atom has an additional neutron, significantly displaces the photocorrosion threshold to ~4× higher values as compared to pure water (Fig. 3a,b). This trend goes well with the 3× lower rate for the radiolysis dissociation products found in Ref. [68] while comparing D$_2$O and H$_2$O.

To further inhibit the photocorrosion and protect the aluminum structures, we use different approaches to passivate the alumina surface and add a supplementary protective layer (Fig. 3c-e). In all these cases, the apertures are filled with 10 mM ascorbic acid at pH 4 which gives the best results in Fig. 3a,b. A first approach uses an extra 10 nm thick silica layer deposited on top of the aluminum film to densify the aluminum oxide layer. As this technique will protect the top aluminum surface but not the aperture walls after FIB milling, the aperture was further treated by 10 minutes oxygen plasma to strengthen the natural oxide layer.[69] This leads to an improvement of the threshold power from 170 µW for the bare aperture to 210 µW for the SiO$_2$ plasma treated aperture. The main advantages of this approach are that it does not involve any wet chemistry, is chemically inert and can be entirely performed during the nanofabrication stage.

A second approach passivates the full aperture surface with a conformal polymer layer to protect from corrosion. We have tested two different polymers: polyvinylphosphonic acid (PVPA)[47] and polydopamine (PDA)[16,30,48]. Both provide a significant extension of the power range where no photocorrosion is observed (Fig. 3c,d), as the threshold power is increased from 170 µW to 270 µW for PVPA and 290 µW for PDA. The slightly better performance of PDA as compared to PVPA can be related to its larger thickness, as our PVPA protocol yields a thickness of about 5 nm,[47] while the thickness for PDA is about 15 nm.[30] If one compares to the case of a bare aperture filled with pure water at pH 7 (threshold power 25 µW, Fig. 3a,b), the optimization of the filling medium and the supplementary use of a polymer protective layer improves the accessible UV laser power range by more than one order of magnitude, with no visible photocorrosion up to 250 µW. Considering that the aluminum corrosion rate goes with the 6$^{th}$ power of the UV intensity, the 10× improvement in the optical power range then translates into a $10^6×$ reduction of the aluminum corrosion rate. To confirm this impressive value, we expose PVPA-protected apertures for up to 15 minutes to 200 µW focused UV light. The SEM images show no visible sign of photocorrosion (Fig. 3e) while under these conditions an untreated water-filled aperture would be almost entirely dissolved over a µm² area (Fig. 2b).

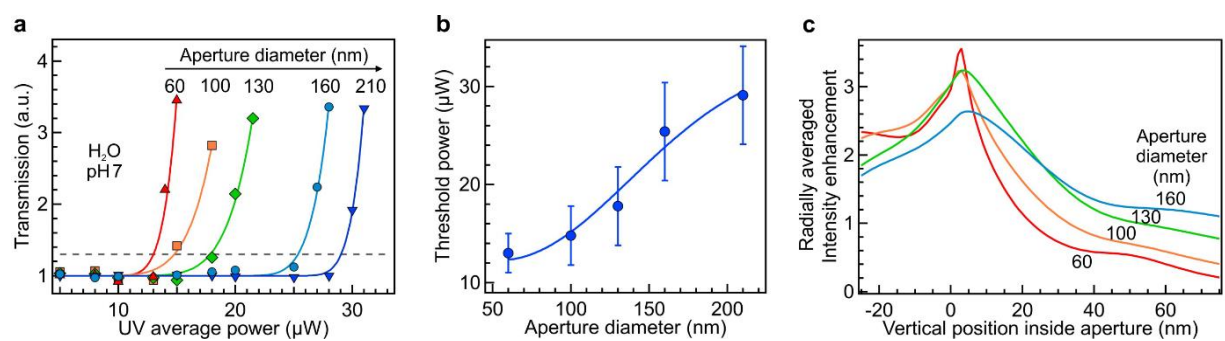

**Figure 4.** Aperture diameter influence on the UV photocorrosion of aluminum. (a) Normalized white light transmission through single apertures of different diameters with increasing UV powers. The apertures are filled with pure water at pH 7 as in Fig. 2. Each data point is taken after a 2 min integration time. The lines are numerical fits with a 6$^{th}$ power law. (b) Evolution of the threshold power deduced from the data in (a) as a function of the aperture diameter. (c) Finite difference time domain (FDTD) numerical simulation of the 266 nm intensity enhancement profiles along the vertical Z axis for different aperture diameters. The profiles show the average intensity enhancement over the aperture XY surface.

So far we have only considered apertures of fixed 160 nm or 1 μm diameter. To probe the influence of the aperture diameter, Fig. 4a shows the normalized transmission recorded for different aperture diameters after increasing UV laser intensities. Since different aperture diameters lead to different transmission levels, we normalize each measured value by the reference transmission in the absence of UV light to obtain a dimensionless parameter allowing to directly compare between different diameters. The dependence with the laser intensity is again clearly non-linear for all the different diameters. From these datasets, we determine for each diameter the threshold power following the same definition as in Fig. 3. As it is already evident from Fig. 4a, we find that the threshold power decreases with the aperture diameter (Fig. 4b). For the smallest 60 nm diameter aperture, only 12 μW of average UV power are enough to initiate the photocorrosion and change the transmission. This strikingly low value even for relatively long 70 ps pulses show that the aluminum photocorrosion process must be considered very seriously for UV plasmonic applications in water environments.

The smallest apertures feature a larger surface to volume ratio. This should benefit to promote their resistance to corrosion as a smaller amount of water is illuminated and a comparatively larger surface of protective alumina layer is present (as compared to larger apertures). However, our observations show a lower power threshold for the smallest apertures, indicating a lower resistance to corrosion that bigger structures. To explain the trend observed experimentally, we have to consider also the influence of the plasmonic local intensity enhancement leading to higher optical fluences inside the nanoaperture. As the corrosion process follows a strong non-linear dependence with the UV optical power, any plasmonic enhancement of the intensity around the aperture will further reduce the threshold power initiating the photocorrosion. We check the influence of the plasmonic intensity enhancement by performing numerical simulations of the 266 nm field intensity distribution inside the nanoaperture. To simplify the presentation of the results, we average the distribution along the horizontal cross-cut surface of the aperture and plot the dependence along the vertical axis (Fig. 4c). For the largest apertures, the UV light essentially propagates through them and small intensity gains are observed. However, when the diameter goes below the 115 nm cut-of diameter for the 266 nm wavelength,[50] the intensity decays evanescently inside the aperture. Simultaneously, the average intensity increases as the diameter is reduced.[70] This intensity enhancement plays a central role in lowering the threshold power defining the occurrence of corrosion for the smallest apertures. It also indicates that appropriate photocorrosion protection strategies are required (Fig. 3) to allow the use of plasmonic nanostructures with high local intensity enhancement factors.

**Conclusion**

Despite the natural alumina layer passivating the aluminum surface, we have found that 266 nm UV light can lead to a dramatic photocorrosion of aluminum in aqueous solutions, even at low average powers in the tens of µW range and within short exposure times of a few tens of seconds. Preventing this effect is a major issue to enable plasmonic sensing and catalysis in the UV range where most organic molecules feature strong absorption bands. The nonlinear two-photon absorption of UV light by water leads to the ionization and dissociation of water molecules and the subsequent production of hydroxyl radicals. These reactive radicals are then the major source causing the aluminum pitting corrosion to occur at the junction between metal grains where the oxide layer is weaker. Adding hydroxyl radical scavengers to the medium filling the aperture and lowering the pH significantly improves the photocorrosion resistance. Additionally, PVPA and PDA polymers can be used to passivate the metal surface and further prevent the corrosion, providing a ten-fold increase in the available UV power range where no photocorrosion is observed. The combination of reactive oxygen species scavengers with conformal protective polymer layers is the key to enable UV plasmonics in aqueous solutions.

**Experimental Section**

*Zero-mode waveguide fabrication*

Cleaned microscope quartz coverslips are coated with a 50 nm-thick layer of aluminum deposited by electron-beam evaporation (Bühler Syrus Pro 710). The chamber pressure during the deposition is maintained below $10^{-6}$ mbar and the deposition rate is 10 nm/s in order to ensure the best plasmonic response for the aluminum layer.[10,71] Individual nanoapertures are then milled using gallium-based focused ion beam (FEI dual beam DB235 Strata) with 30 keV energy and 10 pA current.[72]

*Experimental setup*

The optical microscope is based on a home built confocal setup with a pulsed picosecond 266 nm laser excitation (Picoquant LDH-P-FA-266 laser, 70 ps pulse duration, 80 MHz repetition rate). The laser beam is spatially filtered with a 50 µm pinhole and reflected by a dichroic mirror (Semrock FF310-Di01-25-D) towards the microscope body. A Zeiss Ultrafluar 40x, 0.6 NA glycerol immersion objective focuses the UV laser beam on an individual nanoaperture milled on a 50 nm aluminum layer. The laser spot at the microscope focus has a nearly Gaussian shape with 250 nm full width at half maximum (FWHM). The same microscope objective also collects the light transmitted through the aperture. For the transmission measurements, the light source is the microLED illuminator (Zeiss 423053-9071-000). The detection channel is equipped with a 50 µm pinhole conjugated to the microscope focus for spatial filtering and background noise rejection. A long pass filter (Semrock FF01-300/LP-25) and a bandpass filter (Semrock FF01-375/110-25) further reject the backscattered laser light. A photomultiplier tube (Picoquant PMA 175) connected to a photon counting module (Picoquant Picoharp 300) records the transmitted intensity in the 310 to 410 nm spectral range.

*Surface passivation*

Chemicals are used as received from Sigma Aldrich without further purification. Before passivating the surface with PVPA or PDA polymers, the nanoaperture samples are rinsed with water and isopropanol and then exposed to oxygen plasma for 5 minutes to remove any remaining organic residues. For polyvinylphosphonic acid (PVPA, Sigma Aldrich) passivation, the sample is placed in 2.8 % m/v PVPA solution in water preliminary heated to 90 °C and left for 30 minutes to cover the surface. Then, the sample is removed from the PVPA solution and rinsed with Milli-Q water to wash out the free PVPA residues. Finally, the nanoapertures are annealed at 80 °C for 10 minutes in dry atmosphere. For polydopamine (PDA) passivation, the nanoaperture sample is immersed in a freshly prepared 2 mg/ml dopamine hydrochloride (Sigma Aldrich) solution in TRIS buffer (10mM, pH 8.5, Sigma Aldrich).[30] The sample is kept in the solution for 6 hours at room temperature. During this process dopamine polymerizes and passivates the nanoaperture sample surface.[48] After the passivation is complete, the nanoapertures are rinsed with Milli-Q water and dried with a flow of synthetic air. According to Ref. [30], the PDA coating thickness should reach 15.4 nm as a result of this procedure.

**Acknowledgments**

The authors thank Robert Pansu for stimulating discussions. This project has received funding from the European Research Council (ERC) under the European Union's Horizon 2020 research and innovation programme (grant agreement No 723241).